\documentclass{pasj00}

\begin{document}
\SetRunningHead{Author(s) in page-head}{Running Head}
\Received{2000/12/31}
\Accepted{2001/01/01}

\title{BRIDGE: A Direct-tree Hybrid $N$-body Algorithm for Fully
Self-consistent Simulations of Star Clusters and their Parent Galaxies}

\author{Michiko \textsc{Fujii}}
\affil{Department of Astronomy, Graduate School of Science, The
University of Tokyo, 7-3-1 Hongo, Bunkyo, Tokyo 113-0033}
\email{fujii@astron.s.u-tokyo.ac.jp}

\author{Masaki \textsc{Iwasawa}, Yoko \textsc{Funato}}
\affil{Department of General System Studies, College of Arts and
 Sciences, The University of Tokyo, 3-8-1 Komaba, Meguro, Tokyo 153-8902}
\email{iwasawa@margaux.astron.s.u-tokyo.ac.jp, funato@artcompsci.org}

\and

\author{Junichiro \textsc{Makino}}
\affil{Division of Theoretical Astronomy, National Astronomical Observatory
 of Japan, 2-21-1 Osawa, Mitaka, Tokyo, 181-8588}
\email{makino@cfca.jp}

%


\maketitle

\begin{abstract}
We developed a new direct-tree hybrid $N$-body algorithm for fully
self-consistent $N$-body simulations of star clusters in their
parent galaxies.
In such simulations, star clusters need high accuracy,
while galaxies need a fast scheme because of the large number of the
particles required to model it.
In our new algorithm, the internal motion of the star cluster is
calculated accurately using the direct Hermite scheme with
individual timesteps and all other motions are calculated using the tree
code with second-order leapfrog integrator.
The direct and tree schemes are combined using an extension of
the mixed variable symplectic (MVS) scheme.
Thus, the Hamiltonian corresponding to everything other than the
internal motion of the star cluster is integrated with the leapfrog,
which is symplectic.
Using this algorithm, we performed fully self-consistent $N$-body
simulations of star clusters in their parent galaxy.
The internal and orbital evolutions of the star cluster agreed well with
those obtained using the direct scheme.
We also performed fully self-consistent $N$-body simulation for
large-$N$ models ($N=2\times 10^6$).
In this case, the calculation speed was seven times faster than what
would be if the direct scheme was used.
\end{abstract}

\section{Introduction}

Very young and massive stars have been
found in the central parsec of the Galaxy \citep{Kr95,Pa01,Pa06}.
These stars are very young ($<10$ Myr) \citep{Pa01,Ghez03} and lie on
two disks \citep{Pa06}.
The disks rotate around the central black hole (BH) and are at a large
angle with each other. One disk rotates clockwise in projection, the
other counterclockwise, although the existance of the counterclockwise
disk is controversial \citep{Lu06}.
These disks are coeval to within 1 Myr. 

However, in situ formation of these stars seems unlikely since the
strong tidal field of the central super massive black hole prevents the
gas density from reaching
the Roche value high enough for star formation through self gravity. 
To solve this problem, two possibilities have been
suggested: (1) star formation in situ in a dense gaseous disk around
Sgr A*, or (2) infall of a young star cluster.

The disk model was proposed by \citet{LB03}.
A dense gaseous disk is formed, when a infalling molecular cloud is
tidally disrupted by the central BH. If the density exceeds
$M/2\pi R^3$ at radius $R$, where $M$ is the mass of the central BH, 
the disk becomes unstable with respect to its
own gravity and starts to fragment. 
The fragments collapse and star formation occurs. 
However, this model is problematic. Observations have shown that two
disks are at large angles with respect to each other and these stars
on the disks formed almost at the same time. In this in
situ formation scenario, two disks at large angles must have existed
within 1 Myr, unless one disk changed its orientation through, for
example, external perturbation.

The star cluster inspiral model was proposed by \citet{Gerhard01}.
A star cluster is formed at a distance of tens of parsecs from the
Galactic center (GC)
and spiraled into the GC due to dynamical friction
before being disrupted by the tidal field of the central black hole.
In the central 30 pc, two young dense star clusters, the Arches and
Quintuplet clusters, are observed \citep{Na95,Fi04}.
This model can naturally explain the existence of two stellar disks.
Numerical simulations of this model have been performed 
(\cite{PZ03}, hereafter PZ03; \cite{KM03,GR05}).
These simulations showed that it took too long for the star cluster to
sink to the central parsec unless it was very massive
($>10^6M_{\odot}$) or the initial position of the star cluster is very
near from the GC (2pc or less).

In these simulations (PZ03; G\"{u}rkan \& Rasio 2005), the dynamical
friction from the Galaxy on the star cluster was calculated analytically
using the dynamical friction formula \citep{Ch43}.
However, it is not clear whether this analytical treatment of the
orbital evolution is correct in the case of star clusters.
For rigid objects, the orbital evolutions obtained using the analytical
treatment agree well with those obtained using $N$-body simulations,
if $\ln \Lambda$ is determined correctly taking into account the distance
from the galactic center and the size of the object \citep{H03, Sp03}.
However, our fully self-consistent $N$-body simulation of a satellite
galaxy within its parent galaxy showed that the orbital decay of the
satellite is much faster than those calculated analytically using the
dynamical friction formula, even if the correct value of $\ln \Lambda$
is used \citep{Fj06}.
This difference is caused by the enhancement of the dynamical friction
by particles escaped from the satellite. 
In the case of star clusters, the same enhancement should occur.
Thus, a fully self-consistent $N$-body simulation is necessary to obtain
correct results for the orbital evolution of star clusters.

Such a fully self-consistent $N$-body simulation has not been performed
because it was impossible with existing numerical methods.
There are two classes of numerical methods for $N$-body simulations. 
One is the direct summation, suitable for collisional systems, and the
other is schemes for collisionless systems such as tree, SCF, and
particle-mesh schemes. 
Neither of them can handle such multi-scale systems like star clusters
within galaxies.

The direct summation is the simplest and the most accurate way to
calculate forces;
for a particle we calculate $N-1$ inter-particle forces and sum them up.
Thus the cost to integrate all $N$ particles is $O(N^2)$.
The combination of direct summation, Hermite integrator and individual
timesteps is widely used. 
It can achieve very high accuracy.
Hence, this method is used for simulations of collisional systems
such as star clusters, planetary formation, and galactic nuclei with
central BHs. However, we cannot use this method for galaxies because its
cost is too high.

The tree algorithm \citep{BH86,M04} is an approximate algorithm to
calculate forces.
It can achieve $O(N\log N)$ scaling.
This method is often used with 2nd-order leapfrog integrator.
This method is very powerful for systems which need many particles and
in which particles have similar timescales, such as galaxies or large
scale structures.
Tree code is difficult to combine with individual timesteps and high
order integration schemes. 
Therefore, the tree algorithm is not suitable for collisional systems.
Other schemes for collisionless systems have similar proper times.

Fully self-consistent $N$-body simulations of star clusters which
orbit in their parent galaxy require (1) accurate time integration
for star clusters and (2) fast time integration for the parent galaxy.
Neither of direct or tree (or collisionless) methods, however, can
satisfy these requirements.
This is the reason why fully self-consistent $N$-body simulations of
star clusters within galaxies have never been performed.

A way to solve this problem is an algorithm based on tree code
with individual timesteps \citep{HK89,MA93}. 
This approach, however, has problems with the accuracies
and/or the costs.
Furthermore, it is difficult to use GRAPE with such a method.

We developed a new algorithm which is an extension of the mixed
variable symplectic (MVS) method developed for long-term integration of
planetary systems \citep{WH91,KYN91}.
In our new scheme, star clusters are integrated using direct summation
and Hermite scheme with individual timesteps, while galaxies are
integrated using the tree code and leapfrog scheme with shared timestep.
We combine them by embedding direct scheme into the tree code.
Using this new algorithm, we have made it possible to perform fully
self-consistent $N$-body simulations of star clusters which orbit in their
parent galaxies with very accurate time integration for star clusters and
fast time integration for the parent galaxy.

We describe our new algorithm in section 3 after a description of the
MVS method in section 2. In section 4, we show the results of test
simulations and performance of our new algorithm. 
We summarize in section 5.

\section{The Mixed Variable Symplectic Method}
The MVS integrator were introduced by \citet{WH91} and
by Kinoshita, Yoshida, \& Nakai (1991). It is now widely used for
long-term integrations of planetary systems.
In the case of planetary systems, their Hamiltonian can be divided into
Kepler motions and interaction between planets. 
The MVS integrator suppresses the error of the motion due to the
numerical integration of the solar potential since it integrates the
Kepler motions analytically.

Let us first describe briefly a symplectic integrator, the leapfrog scheme.
The Hamilton equation is rewritten in terms of a Poisson bracket operator as
\begin{eqnarray}
\frac{df}{dt} = \{f,H\},
\label{eq:df}
\end{eqnarray}
where $f$ is a function of $t$. 
If we introduce a differential operator $D_H$ defined as $D_H f$:$=\{f,H\}$,
the formal solution of equation (\ref{eq:df}) is written as
\begin{eqnarray}
f(t) = e^{tD_H} f(0).
\end{eqnarray}
An integration algorithm can be thought of as an approximate expression
of this operator.

As an example, we describe a second-order leapfrog integrator.
The Hamiltonian for an $N$-body system is written as
\begin{eqnarray}
H &=& \sum^{N}_{i}\frac{p_i^2}{2m_i} - \sum^{N}_{i<j}\frac{Gm_i m_j}{r_{ij}}.
\end{eqnarray}
If we define 
\begin{eqnarray}
H_A &=& - \sum^{N}_{i<j}\frac{Gm_i m_j}{r_{ij}} \\
H_B &=& \sum^{N}_{i}\frac{p_i^2}{2m_i},
\end{eqnarray}
then we can express the the formal solution, the time evolution from $t$
to $t+\Delta t$, as
\begin{eqnarray}
f(t+\Delta t) = e^{\Delta t(A+B)} f(t),
\end{eqnarray}
where $A$:$=D_{H_A}$ and $B$:$=D_{H_B}$.
This operator can be rewritten by the Taylor series as
\begin{eqnarray}
e^{\Delta t(A+B)} = \prod^{k}_{i=1} e^{a_i \Delta t A}e^{b_i \Delta t B} 
+ O(\Delta t ^{n+1}),
\label{eq:Taylor}
\end{eqnarray}
where $(a_i, b_i)$ $(i=1, 2, \dots ,k)$ is a set of real numbers and
$n$ is a given integer, which corresponds to the order of integrator. 
By neglecting the $O(\Delta t^{n+1})$ then, we obtain a mapping from
$f(t)$ to $f'(t+\Delta t)$ as
\begin{eqnarray}
f'(t+\Delta t) = \prod^{k}_{i=1} e^{a_i \Delta t A}e^{b_i \Delta t B} 
f(t).
\end{eqnarray}
This mapping is symplectic because it is just a product of symplectic
mappings.
This is an $n$-th order symplectic integrator.
We can achieve $n=2$ with k=2, with the choice of the coefficients
$a_1=a_2=1/2$ and $b_1=1, b_2=0$. Now, equation (\ref{eq:Taylor}) is
reduced to
\begin{eqnarray}
e^{\Delta t (A+B)}=e^{\frac{1}{2}\Delta tA}e^{\Delta tB}
e^{\frac{1}{2}\Delta tA} + O(\Delta t ^3).
\end{eqnarray}
Therefore the time evolution is expressed as 
\begin{eqnarray}
f'(t+\Delta t) = e^{\frac{1}{2}\Delta tA}e^{\Delta tB}
e^{\frac{1}{2}\Delta tA} f(t).
\label{eq:time_evolution}
\end{eqnarray}
This is the second-order leapfrog scheme, which is rewritten as
\begin{eqnarray}
\boldsymbol{v}_{\frac{1}{2}} &=& \boldsymbol{v}_{0} + \frac{1}{2}\ \Delta t\ \boldsymbol{a}_{0},
\label{eq:lpv}\\
\boldsymbol{x}_{1} &=& \boldsymbol{x}_{0} + \Delta t\ \boldsymbol{v}_{\frac{1}{2}},
\label{eq:lcx}\\
\boldsymbol{v}_{1} &=& \boldsymbol{v}_{0} + \frac{1}{2}\ \Delta t\ \boldsymbol{a}_{1},
\label{eq:lcv}
\end{eqnarray}
where subscripts, 0, $\frac{1}{2}$, 1, correspond to values at
$t, t+\frac{1}{2}\Delta t, t+\Delta t$, respectively.

The procedure of leapfrog scheme is as follows.
\begin{enumerate}
\item Calculate the acceleration at a time, $t$, and update the velocity 
      [eq. (\ref{eq:lpv})].
\item Update positions using new velocity $\boldsymbol{v}_{\frac{1}{2}}$ 
      [eq. (\ref{eq:lcx})].
\item Calculate the acceleration at $t+\Delta t$ using the new positions, 
      $\boldsymbol{x}_1$, and update velocity [eq. (\ref{eq:lcv})].
\item Repeat 1-3.
\end{enumerate}

Now, we explain an MVS integrator.
The Hamiltonian for a planetary system can be expressed as
\begin{eqnarray}
H = H_{\rm Kep} + H_{\rm Int},
\end{eqnarray}
where $H_{\rm Kep}$ is the kinetic energy plus solar potential and
$H_{\rm Int}$ is the interaction energy between planets.
If we define
\begin{eqnarray}
H_A = H_{\rm Int},\  H_B = H_{\rm Kep},
\end{eqnarray}
equation (\ref{eq:time_evolution}) becomes
\begin{eqnarray}
f'(t+\Delta t)=e^{\frac{1}{2}\Delta tI}e^{\Delta tK}
e^{\frac{1}{2}\Delta tI} f(t).
\end{eqnarray}
Here $I$:=$D_{H_{\rm Int}}$ and $K$:$=D_{H_{\rm Kep}}$. Note that 
$e^{\Delta t K}$ generates motions of planets along unperturbed
Kepler orbits, while $e^{\Delta t I}$ generates changes of momenta due to
planet-planet interactions. 
This changes of momenta are called ``velocity kicks.''

The difference from the usual leapfrog integrator is that 
$e^{\Delta t K}$ is given analytically by Kepler motion.
Therefore the MVS method is expressed as
\begin{eqnarray}
\boldsymbol{v}'_{\frac{1}{2}} &=& \boldsymbol{v}_{0} + \frac{1}{2}\ \Delta t\ \boldsymbol{a}_{\rm
 Int, 0},
\label{eq:MVS1}\\
\boldsymbol{x}_{0} &\rightarrow& ({\rm Kepler\ motion}) \rightarrow \boldsymbol{x}_{1},
\label{eq:MVS2-1} \\
\boldsymbol{v}_{\frac{1}{2}} &\rightarrow& ({\rm Kepler\ motion}) \rightarrow \boldsymbol{v}'_{\frac{1}{2}},\label{eq:MVS2-2} \\
\boldsymbol{v}_{1} &=& \boldsymbol{v}'_{\frac{1}{2}} + \frac{1}{2}\ \Delta t\ \boldsymbol{a}_{\rm Int, 1}.
\label{eq:MVS3}
\end{eqnarray}

The integration proceeds as follows:
\begin{enumerate}
\item Calculate the accelerations of planets due to gravitational
      interactions between planets, 
      $\boldsymbol{a}_{\rm Int, 0}$,
      at time $t$ and change velocities by giving the velocity kicks 
      [eq.(\ref{eq:MVS1})].
\item Update the positions and the velocities by $\Delta t$ along its
      osculating Kepler orbit analytically 
      [eq.(\ref{eq:MVS2-1}) and (\ref{eq:MVS2-2})].
\item Calculate $\boldsymbol{a}_{\rm Int, 0}$
      at $t+\Delta t$ and change velocities by giving the velocity kicks 
      [eq.(\ref{eq:MVS3})].
\item Repeat 1-3.
\end{enumerate}

MVS is a very powerful algorithm for long-term integration of planetary
systems.
In general, the integration errors are $O(\Delta t ^n)$, where $n$ is the
order of the integrator. With a MVS integrator, if $H_{\rm Int}$ is
$O(\epsilon)$ of $H_{\rm Kepler}$,
the integration errors are only $O(\epsilon \Delta t^n)$. 
This $\epsilon$ is of the order of the planetary mass in the unit of
solar mass and usually is very small.
As a result, the error become much smaller than that of usual symplectic
methods.

\section{The New Hybrid Scheme}

Now we consider simulations of systems consisting of a star cluster
and its parent galaxy. For such a simulation, our new scheme should provide:
\begin{enumerate}
\item high accuracy for star clusters
\item fast integrator for galaxies
\item fully self-consistent treatment of the total system.
\end{enumerate}
To achieve these goals, we constructed a new scheme which is a combination
of the direct and the tree scheme.
In our new scheme, the internal interactions of star clusters are
calculated with high accuracy using the direct and Hermite
scheme, while all other interactions (galaxy-galaxy, galaxy-star
cluster) are calculated with the tree algorithm.
We combine these two methods by extending the idea of the MVS.

In the MVS scheme, the Hamiltonian is divided into the kinetic energy
plus solar potential and the interaction energy between planets.
In our hybrid scheme, we separate the Hamiltonian as
\begin{eqnarray}
H &=& H_{\rm \alpha} + H_{\rm \beta},\\
H_{\rm \alpha} &=& -\sum ^{N_{\rm G}}_{i<j} \frac{Gm_{{\rm G},i}m_{{\rm
 G},j}}{r_{ij}}
 - \sum ^{N_{\rm G}}_{i=1} \sum^{N_{\rm SC}}_{j=1} \frac{Gm_{{\rm G},i}
 m_{{\rm SC},j}}{r_{ij}},\\
H_{\rm \beta} &=& \sum^{N_{\rm G}}_{i=1} \frac{p_{{\rm G},i}^2}{2m_{{\rm
 G},i}}
 + \sum^{N_{\rm SC}}_{i=1} \frac{p_{{\rm SC},i}^2}{2m_{{\rm SC},i}}
 - \sum ^{N_{\rm SC}}_{i<j} \frac{Gm_{{\rm SC},i} m_{{\rm SC},j}}{r_{ij}},
\label{eq:H_beta}
\end{eqnarray}
where $N_{\rm G}$ and $N_{\rm SC}$ are the number of the galaxy
particles  star cluster particles, respectively, and $H_{\rm \alpha}$ is the
potential energy of the gravitational interactions between galaxy
particles and between galaxy particles and star cluster particles, 
while $H_{\rm \beta}$ is the kinetic energy of
all particles and the potential energy of star cluster particles.
Using the replacement $H_A = H_{\rm \alpha}$ and $H_B = H_{\rm \beta}$, 
we obtain the time evolution as
\begin{eqnarray}
f'(t+\Delta t)=e^{\frac{1}{2}\Delta t\alpha}e^{\Delta t\beta}
e^{\frac{1}{2}\Delta t\alpha} f(t),
\end{eqnarray}
where $\alpha$:=$D_{H_{\rm \alpha}}$, $\beta$:$=D_{H_{\rm \beta}}$.

In our scheme, we integrate star cluster particles and galaxy
particles in different ways.
Let us first discuss star cluster particles. They are integrated in a
way similar to the MVS.
The Keplerian in the MVS corresponds to the second and third terms in
equation (\ref{eq:H_beta}).
Unlike the MVS, however, we cannot solve the Hamiltonian analytically.
Hence, we replace an analytical solution (Kepler motion) in the MVS
by a solution calculated by a higher-order integrator (e.g. the
fourth-order Hermite integrator with individual timesteps).

Thus, the integrator for star clusters is written as
\begin{eqnarray}
\boldsymbol{v}'_{\rm SC,0} &=& \boldsymbol{v}_{\rm SC,0} + \frac{1}{2}\ \Delta
 t\ \boldsymbol{a}_{\rm \{ G\rightarrow SC,0\}},\label{eq:hybrid1}\\
\boldsymbol{x}_{\rm SC, 0} &\rightarrow& ({\rm Hermite\ scheme})\rightarrow \boldsymbol{x}_{\rm SC, 1},\label{eq:hybrid2}\\
\boldsymbol{v}'_{\rm SC,0} &\rightarrow& ({\rm Hermite\ scheme})\rightarrow \boldsymbol{v}'_{\rm SC,1},\label{eq:hybrid3}\\ 
\boldsymbol{v}_{\rm SC,1} &=& \boldsymbol{v}'_{\rm SC,1} + \frac{1}{2}\
 \Delta t\ \boldsymbol{a}_{\rm \{ G\rightarrow SC,1\}},\label{eq:hybrid4}
\end{eqnarray}
where subscripts, SC and G, stand for the star cluster and the galaxy,
subscripts, 0, $\frac{1}{2}$, and 1, indicate times $t_0$,
$t_{\frac{1}{2}}=t_0+\frac{1}{2}\Delta t$, and $t_1=t_0+\Delta t$,
respectively, and 
$\boldsymbol{v}'_{\rm SC,\frac{1}{2}}$ represent a new velocity at
$t_{\frac{1}{2}}$, which have been integrated using the Hermite scheme.

For galaxies, we use the leapfrog integrator expressed as
\begin{eqnarray}
\boldsymbol{v}_{\rm G, \frac{1}{2}} &=& \boldsymbol{v}_{\rm G,0} + \frac{1}{2}\ \Delta
 t\ \boldsymbol{a}_{\{ \rm All\rightarrow G,0\}}, \label{eq:hybrid5}\\
\boldsymbol{x}_{\rm G,1} &=& \boldsymbol{x}_{\rm G,0} + \Delta t\ \boldsymbol{v}_{\rm G, \frac{1}{2}},\label{eq:hybrid6} \\
\boldsymbol{v}_{\rm G,1} &=& \boldsymbol{v}_{\rm G, \frac{1}{2}} + \frac{1}{2}\ \Delta t\ \boldsymbol{a}_{\rm \{All\rightarrow G,1\}},\label{eq:hybrid7}
\end{eqnarray}
where $\boldsymbol{a}_{\rm \{All\rightarrow G\}}$ denotes the
acceleration due to gravitational forces from all
particles (including star cluster particles) to the galaxy particle.
The galaxy particles have longer timescale than that of particles in the
star cluster. 
Therefore, we adopt a second-order leapfrog integrator with shared
timestep and tree algorithm.
This scheme is less accurate than fourth-order Hermite scheme, but much
faster and is symplectic.

\begin{figure}[htbp]
  \begin{center}  
    \FigureFile(80mm,50mm){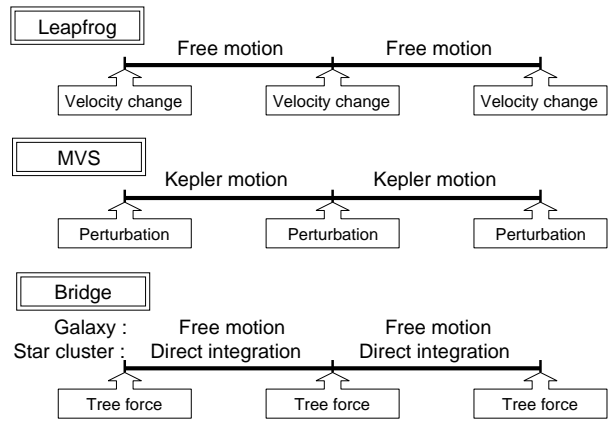}
  \end{center}
  \caption{Procedure of the Bridge scheme.}
  \label{fig:scheme}
\end{figure}

We call our new scheme ``the Bridge scheme'' (Bridge is for Realistic
Interactions in Dense Galactic Environment).
The procedure of the Bridge scheme is summarized in figure
\ref{fig:scheme} and as follows:
\begin{enumerate}
\item Make a tree at $t_0$ and calculate the accelerations from all
      particles on galaxy particles, 
      $\boldsymbol{a}_{\rm \{All \rightarrow G, 0\}}$,
      and from galaxy particles on star cluster particles, 
      $\boldsymbol{a}_{\rm \{G \rightarrow SC, 0\}}$, using the tree.
\item {\bf Star cluster}: give a velocity kick [eq. (\ref{eq:hybrid1})].\\
      {\bf Galaxy}: update velocity [eq. (\ref{eq:hybrid5})].
\item {\bf Star cluster}: integrate positions and the velocities from $t_0$
      to $t_1$ using Hermite scheme with individual timesteps
      [eq. (\ref{eq:hybrid2}) and (\ref{eq:hybrid3})].\\
      {\bf Galaxy}: Update position with leapfrog
      scheme [eq. (\ref{eq:hybrid6})].
\item Make a new tree at $t_1$ and calculate the accelerations from all
      particles on galaxy particles, 
      $\boldsymbol{a}_{\rm \{All \rightarrow G, 1\}}$, 
      and from galaxy particles on star cluster particles, 
      $\boldsymbol{a}_{\rm \{G \rightarrow SC, 1\}}$.
\item {\bf Star cluster}: give a velocity kick [eq. (\ref{eq:hybrid4})].\\
      {\bf Galaxy}: update velocity [eq. (\ref{eq:hybrid7})].
\end{enumerate} 

As shown in equations (\ref{eq:hybrid1}) and (\ref{eq:hybrid5}), the
forces on particles of galaxies and those on particles of star clusters
are calculated differently. The former is from all particles, while the
latter is from particles in the galaxy only. Therefore, we assigned two
values of mass to each tree node. (We used center-of-mass approximation
to forces from tree nodes.) One is total mass of all particles under the
node, and the other is the mass of galaxy particles. To calculate forces
on galaxy particles, we use the mass of all particles, and for star cluster
particles we use the mass of galaxy particles. 

\section{Accuracy and Performance}
\subsection{Comparison with Direct Scheme for Small-$N$ Model}
As a test of the Bridge scheme, we performed a fully self-consistent
$N$-body simulation of a star cluster within a galaxy and compared the
results with those obtained with the fourth-order Hermite scheme.
In this section, we describe the results and the performance of the
Bridge scheme.

We adopted a King model with the non-dimensional central potential
$W_0=9$ as the model of the parent galaxy and with $W_0=7$ as that
of the star cluster.
The system of units is the Heggie unit (Heggie \& Mathieu 1986), where 
the gravitational constant $G$ is 1 and the mass and the binding energy
of the parent galaxy are 1 and 0.25, respectively. The initial position
of the star cluster is at distance 2.5 from the center of the parent
galaxy and the initial velocity is 0.65.
Both the galaxy and the star cluster are expressed as a self-consistent
$N$-body models. The number of particles of the parent galaxy, $N_{\rm G}$,
is $10^5$ and that of the star cluster, $N_{\rm SC}$, is $2\times 10^3$. 
In table \ref{tb:testrun}, we summarize the model parameters and initial
conditions. If we assume the total mass of the star cluster 
$M_{\rm SC}=10^5 \MO$ and the unit length is 10 pc, the total mass of
the Galaxy $M_{\rm G}=10^7 \MO$ and the unit time and velocity are
0.15 Myr and 66 km s$^{-1}$, respectively. These values would correspond
to the central region of a galaxy somewhat smaller than our galaxy.

The potential is softened using the usual Plummer softening.
The softening length for the gravitational interactions between star
cluster particles is
$\epsilon _{\rm SC} = 2.0 \times 10^{-4}=0.1 \times 4/N_{\rm SC}$,
and that between others (i.e. between galaxy particles and for
galaxy particles and star cluster particles) is
$\epsilon _{\rm G} = 6.25 \times 10^{-3}$.

We used the opening angle $\theta = 0.75$ with the center-of-mass
(dipole-accurate) approximation. 
The maximum group size for GRAPE calculation (Makino 1991) is 8192. 
For the leapfrog integrator, we adopted the stepsizes of 
$\Delta t=1/128$ and 1/256.
The maximum timestep for the Hermite scheme with individual timesteps
is equal to the timestep of the tree. All particles synchronize at each
timestep for tree. Within these steps, star cluster particles are
integrated the Hermite scheme with individual timesteps
\citep{MA92}. For the timestep criterion, we adopted the standard
formula, which is given in \citet{MA92}.
These parameters are summarized in table \ref{tb:param}.

The simulations are performed on GRAPE-6 \citep{M03} for
runs Direct 1 and 2, GRAPE-6A \citep{Fk05} for runs Bridge 1
and 2.
We summarize them in table \ref{tb:runs}.
Total energy was conserved to better than 0.06\% 
when $\Delta t=1/256$, 0.2\% $\Delta t=1/128$ with the Bridge scheme,
and 0.006\% with the Hermite scheme.

To compare the results, we followed the time evolution of the position, 
bound mass, core radius, and core density of the star cluster. 
The bound mass and orbit are calculated in the same way as in \citet{Fj06}.
We defined the core radius and the core density using the formula
proposed by \citet{CH85}.

\begin{table}[htbp]
\begin{center}
\caption{Model Parameters of Testmodel \label{tb:testrun}}  
\begin{tabular}{lcc}
\hline \hline
Parameters & Galaxy & Star cluster \\ \hline
Galactic halo & King 9 & King 7 \\ 
Total mass & 1.0 & 0.01  \\ 
Binding energy & 0.25 & $2.5 \times 10^{-4}$ \\ 
Half-mass radius & $9.8\times 10^{-1}$ & $8.1\times 10^{-2}$\\
$N$    & $10^5$& $2\times 10^3$ \\ \hline
Initial position & & ( 2.5, 0, 0) \\ 
Initial velocity & &( 0, 0.65, 0) \\ \hline
\end{tabular}
\end{center}
\end{table}

\begin{table}[htbp]
\begin{center}
\caption{Parameters for $N$-body Simulation\label{tb:param}}  
\begin{tabular}{lcc}
\hline \hline
Parameters & Value \\ \hline
$\epsilon _{\rm G}$ &  $6.25 \times 10^{-3}$ \\
$\epsilon _{\rm SC}$ &  $2.0 \times 10^{-4}$ \\ \hline
$\theta$ & 0.75\\
$n_{\rm crit}$ & 8192\\
$\Delta t$ & 1/128, 1/256\\ \hline
\end{tabular}
\end{center}
\end{table}

\begin{table}
\begin{center}
\caption{Runs\label{tb:runs}}  
\begin{tabular}{lcccc}
\hline \hline
Runs & methods & seed & stepsize &run time (h)\\
\hline
Direct 1 & direct & 1 & 1/256 & 34\\
Direct 2 & direct & 2 & 1/256 & 34\\
Bridge 1 & hybrid & 1 & 1/128 & 10\\
Bridge 2 & hybrid & 2 & 1/256 & 19\\
\hline
\end{tabular}
\end{center}
\end{table}

\begin{figure}[htbp]
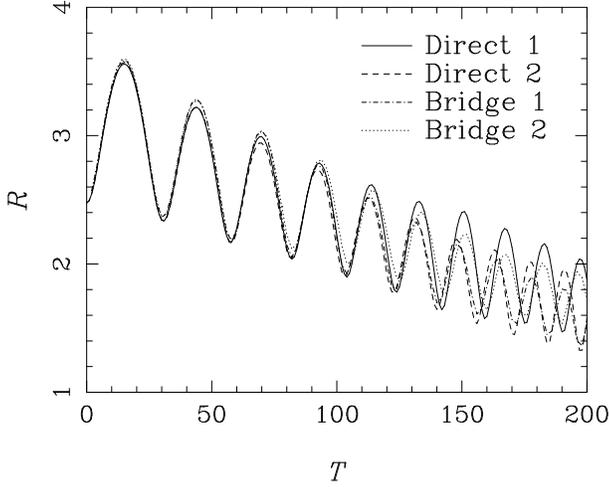

\begin{center}
 \FigureFile(80mm,50mm){figure2.eps}
\end{center}

 \caption{Distance of the star cluster from the GC plotted as a
 function of time. Solid and dashed curves show the results of the runs
 in which all particles calculated with the direct (Hermite) scheme. 
 Dashed-doted and dotted curves show those with the Bridge scheme.}
  \label{fig:radius_test}
\end{figure}

\begin{figure}[htbp]
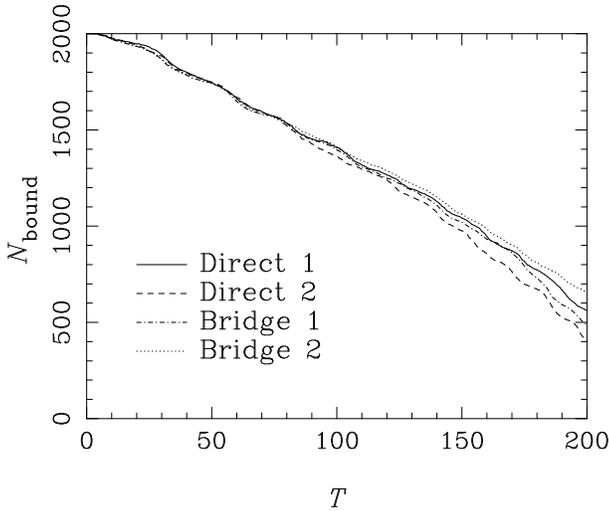

 \begin{center}  
  \FigureFile(80mm,50mm){figure3.eps}
 \end{center}
  \caption{Bound mass of the star cluster plotted as a function of time.
 Curves have the same meanings as in figure \ref{fig:radius_test}.}
 \label{fig:mass_test}
\end{figure}

\begin{figure}[htbp]
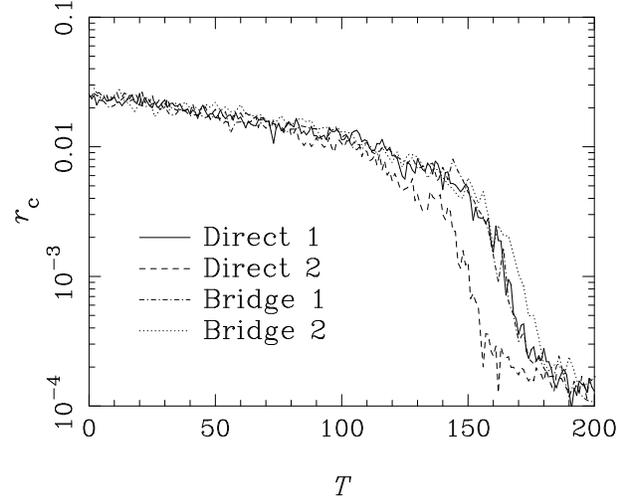

  \begin{center}  
    \FigureFile(80mm,50mm){figure4.eps}
  \end{center}
  \caption{Core radius, $r_{\rm c}$, plotted as a function of time.
 Curves have the same meanings as in figure \ref{fig:radius_test}.}
  \label{fig:core_test}
\end{figure}

\begin{figure}[htbp]
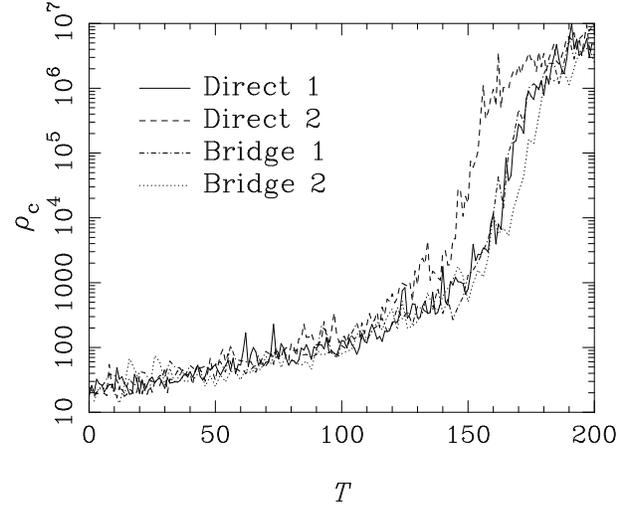

  \begin{center}  
    \FigureFile(80mm,50mm){figure5.eps}
  \end{center}
  \caption{Core density, $\rho _{\rm c}$, as a function of time.
 Curves have the same meanings as in figure \ref{fig:radius_test}.}
  \label{fig:density_test}
\end{figure}

Figures \ref{fig:radius_test} and \ref{fig:mass_test} show the evolution
of the distance from the galactic center and the bound mass of the star
cluster.
Figures \ref{fig:core_test} and \ref{fig:density_test} show the core
radius and the core density of the star cluster.
These results show that the Bridge scheme works very well.
The difference between the results of the Hermite scheme and that of the
Bridge scheme is smaller than run-to-run variations in each method.

Figures \ref{fig:core_test} and \ref{fig:density_test} show that core
collapse occurs at $T=150-180$.
Core collapse occurs at 
\begin{eqnarray}
t_{\rm cc} \simeq ct_{\rm rh},
\label{eq:t_cc}
\end{eqnarray}
where $t_{\rm rh}$ is the star cluster's half-mass relaxation time
\citep{SH71},
\begin{eqnarray}
t_{\rm rh} = 0.14 \frac{r_{\rm h}^{3/2}\ N_{\rm SC}}
{(GM_{\rm SC})^{1/2}\ln \Lambda}.
\label{eq:t_rh}
\end{eqnarray}

Here $r_{\rm h}$, $N_{\rm SC}$, and $M_{\rm SC}$ are the half-mass radius,
the number of particles, and the mass of the star cluster.
We adopted the Coulomb logarithm $\ln \Lambda \simeq \ln(0.1 N_{\rm SC})$.
In isolated star cluster in which all stars has the same mass, 
$c\simeq 15$ \citep{Cohn80}. From these equations, the core collapse time
of our cluster is calculated as  $t_{\rm cc} \simeq 180$.
This value is consistent with the results of our simulation.
Note that in our model we used the scaling of $M_{\rm G}=4E_{\rm G}=1$.

\begin{figure}[htbp]
  \begin{center}  
    \FigureFile(80mm,50mm){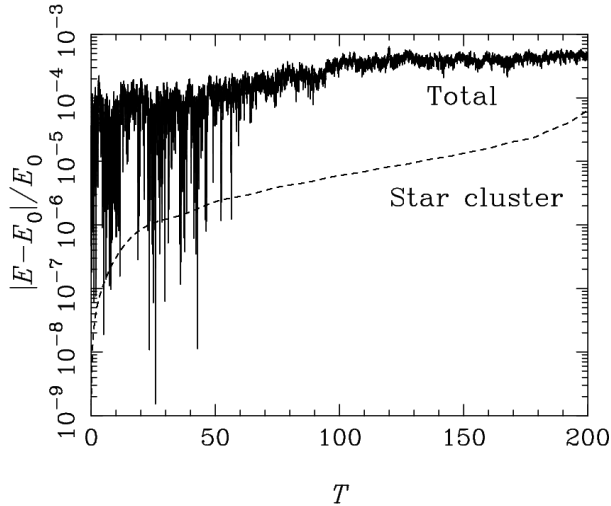}
  \end{center}
  \caption{The energy error of the system, for the hybrid run with
  $\Delta t =1/256$ (Run Bridge 2). Full curve shows the total energy
 error of the system and dashed curve shows the cumulative energy error
 of the star cluster.}
  \label{fig:err_test}
\end{figure}

The total energy error of the system in Bridge 2 run is shown in Figure 
\ref{fig:err_test}. The total energy is conserved very well.
The total energy error depends on the parameters for tree, 
$\Delta t$ and $\theta$.
This is because most of the energy error is generated in the parent
galaxy, which is much larger than the star cluster and has much larger
energy.

To see whether the internal energy of the star cluster is conserved or not,
we  measured the energy error of the internal motion of the star cluster
within each step, $\Delta t$.
The cumulative error of each step is shown in Figure \ref{fig:err_test}.
Note that we plot the energy error of the star cluster relative to the
internal energy of the star cluster, which is 0.1 \% of the total energy
of the system.
Although the error become larger after core collapse occurred, it is
conserved well.

The distributions of the CPU time in runs with $\Delta t=1/128$ and
$\Delta t = 1/256$ are shown in table \ref{tb:performance}. In these
simulations, the cost of the tree part was much larger than that of the
direct part.
The CPU time of the tree part is almost constant throughout the simulation.
In contrast, the cost of the direct part increased after $T \simeq 140$,
because the core density become higher.

\begin{table}[htbp]
\begin{center}
\caption{Distribution of the CPU Time of our Simulations\label{tb:performance}}  
\begin{tabular}{lcc}
\hline \hline
Section of Code & \multicolumn{2}{c}{Percentage of CPU Time (\%)}\\
& $\Delta t = 1/128$ & $\Delta t = 1/256$\\
\hline
Tree & 87.2 & 91.2\\
Direct & 10.7 & 6.6\\
The others & 2.1 & 2.2\\
\hline
\end{tabular}
\end{center}
\end{table}

\subsection{Large-$N$ Models}
We performed simulations of large-$N$ models.
The number of particles is sufficiently large for simulations of star
clusters near the GC ($N_{\rm G}=2\times 10^6$, $N_{\rm SC}=65,536$).
The model of the galaxy represents the central region of the Galaxy from
$\sim 1$ to several pc from the GC.
For the star cluster, we followed the model used in Portegies Zwart et
al. (2003). They  modeled the Arches and Quintuplet star clusters.
The core radius of our model, $r_{\rm core}$, is 0.087 pc.
Using this model, we performed two simulations, in which the star
cluster has circular and eccentric orbits, respectively.
In table \ref{tb:models}, we summarize the model parameters.

\begin{table*}[htbp]
\begin{center}
\caption{Models}
\begin{tabular}{ccccccccc}
\hline
\hline
& King $W_0$& $N$ & $M({\rm M_{\odot}})$ & $r_{\rm c}$ (pc) & $r_{\rm t}$ (pc)\\
\hline
The Galaxy & 10 & $2 \times 10^6 $ & $8.0 \times 10^7$ & 0.66 & 120 \\
Star cluster & 3 & 65536 & $7.9 \times 10^4$ & 0.087 & 0.47 \\
\hline
\end{tabular}
\end{center}
\label{tb:models}
\end{table*}

We performed $N$-body simulations using the Bridge code.
For the tree part, we used the opening
angle $\theta = 0.75$ with the center-of-mass (dipole-accurate)
approximation. The maximum group size for a GRAPE calculation (Makino
1991) is 8192. The stepsize of leapfrog integrator is $\Delta t = 1/512$
(Heggie unit). 
The potential is softened using Plummer softening.
The softening length for gravitational interactions between star cluster
particles, $\epsilon _{\rm SC}$, is $1.0\times 10^{-5}$ pc and 
that for others, $\epsilon _{\rm G}$, is $3.9\times 10^{-2}$ pc. 
We stopped the simulations at $T =0.75 ({\rm Myr}) = 5 ({\rm unit\ time})$.  
These parameters are summarized in table \ref{tb:param2}.
After the core collapse, the structures of the star clusters are
not expressed correctly in our simulations because we use a softened
potential for stars.

\begin{figure}[htbp]
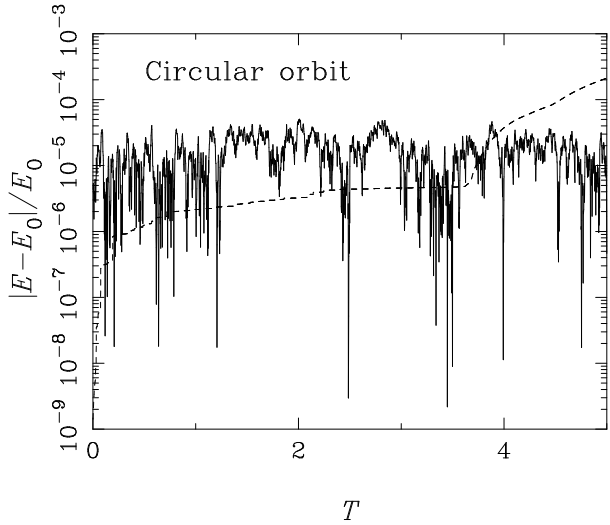

  \begin{center}  
    \FigureFile(80mm,50mm){figure9.eps}
  \end{center}
  \caption{Same as figure \ref{fig:err_test}, but for large-$N$ model
 with circular orbit.}
  \label{fig:err_cir}
\end{figure}

\begin{figure}[htbp]
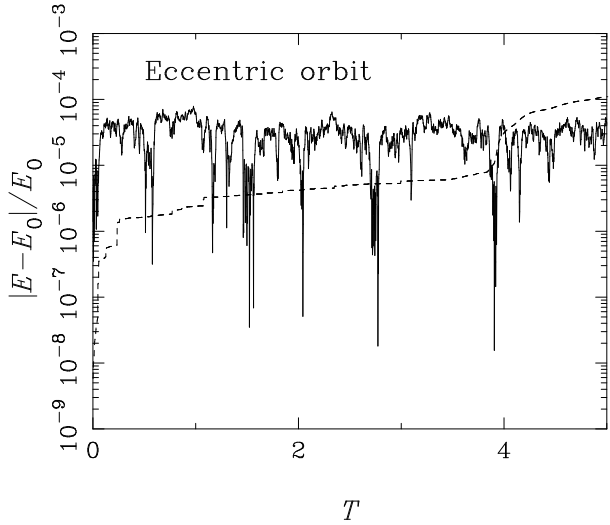

  \begin{center}  
    \FigureFile(80mm,50mm){figure10.eps}
  \end{center}
  \caption{Same as figure \ref{fig:err_test}, but for large-$N$ model
 with eccentric orbit.}
  \label{fig:err_ecc}
\end{figure}

We used GRAPE-6 (Makino et al. 2003) for force calculation.
The total energy was conserved better than $5\times 10^{-5}$ for
the circular orbit (figure \ref{fig:err_cir}) and 
$8\times 10^{-5}$ for the eccentric orbit 
(figure \ref{fig:err_ecc}) throughout the simulations.

\begin{table}[htbp]
\begin{center}
\caption{Parameters for $N$-body Simulation\label{tb:param2}}  
\begin{tabular}{lcc}
\hline \hline
Parameters & Value \\ \hline
$\epsilon _{\rm G}$ &  $3.9 \times 10^{-2}$ (pc)\\
$\epsilon _{\rm SC}$ & $1.0 \times 10^{-5}$ (pc)\\ \hline
$\Delta t$ & $2.9 \times 10^{-4}$ (Myr)\\
      (Heggie unit) &1/512 \\ \hline
$\theta$ & 0.75\\
$n_{\rm crit}$ & 8192\\ \hline
\end{tabular}
\end{center}
\end{table}

\begin{table}[htbp]
\begin{center}
\caption{Initial Conditions}
\begin{tabular}{ccccc}
\hline
\hline
Simulation & Initial position (pc) & Initial velocity (km s$^{-1}$)\\ \hline
Circular & 2 & 130\\ 
Eccentric & 5 & 72 \\
\hline
\end{tabular}
\end{center}
\end{table}

\begin{figure*}[htbp]
  \begin{center}  
    \FigureFile(158mm,238mm){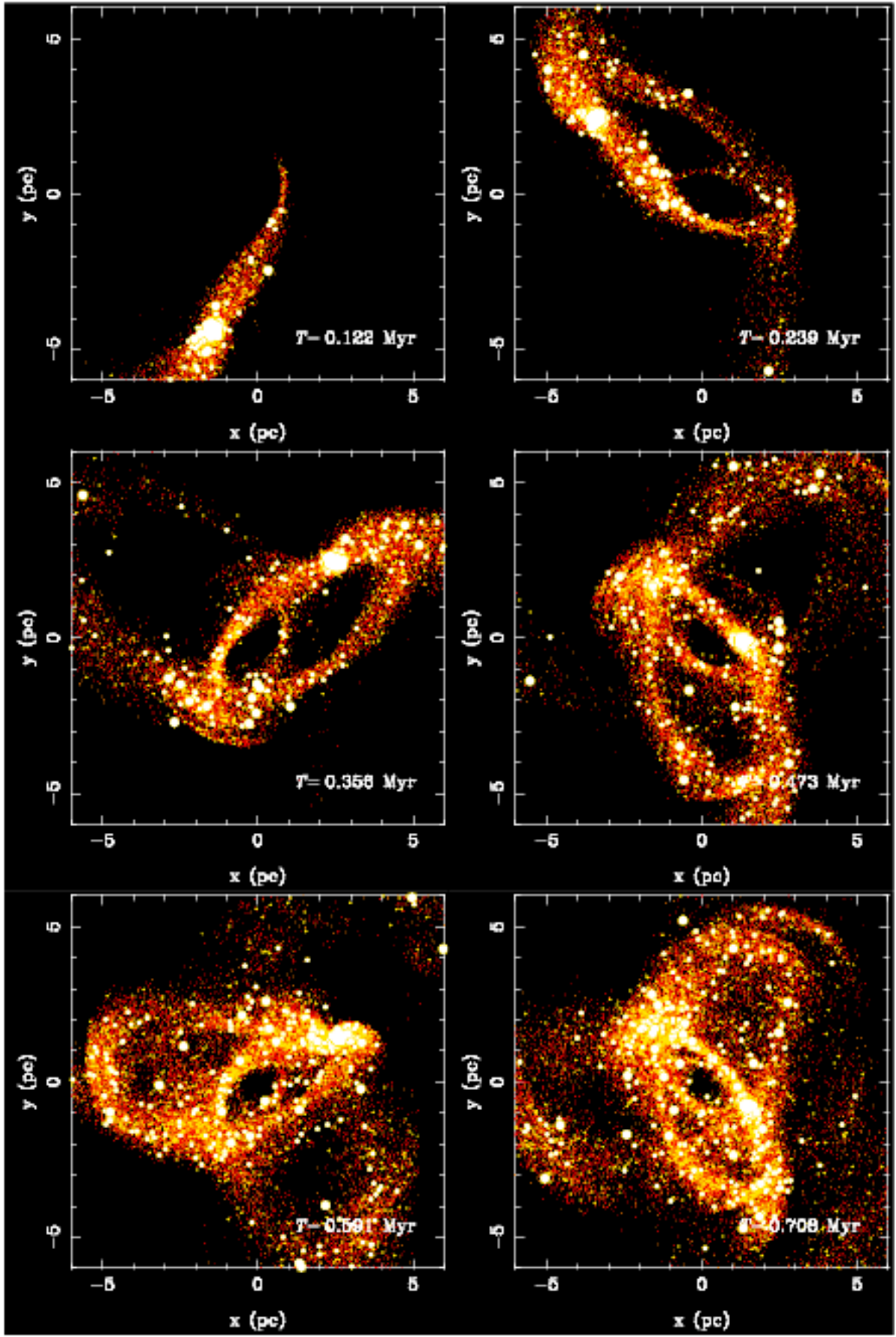}
  \end{center}
 \caption{Snapshots of the star clusters projected onto $x-y$ plane. The
 orbit of the star cluster is eccentric.}
 \label{fig:snapshots}
\end{figure*}

\begin{figure*}[htbp]
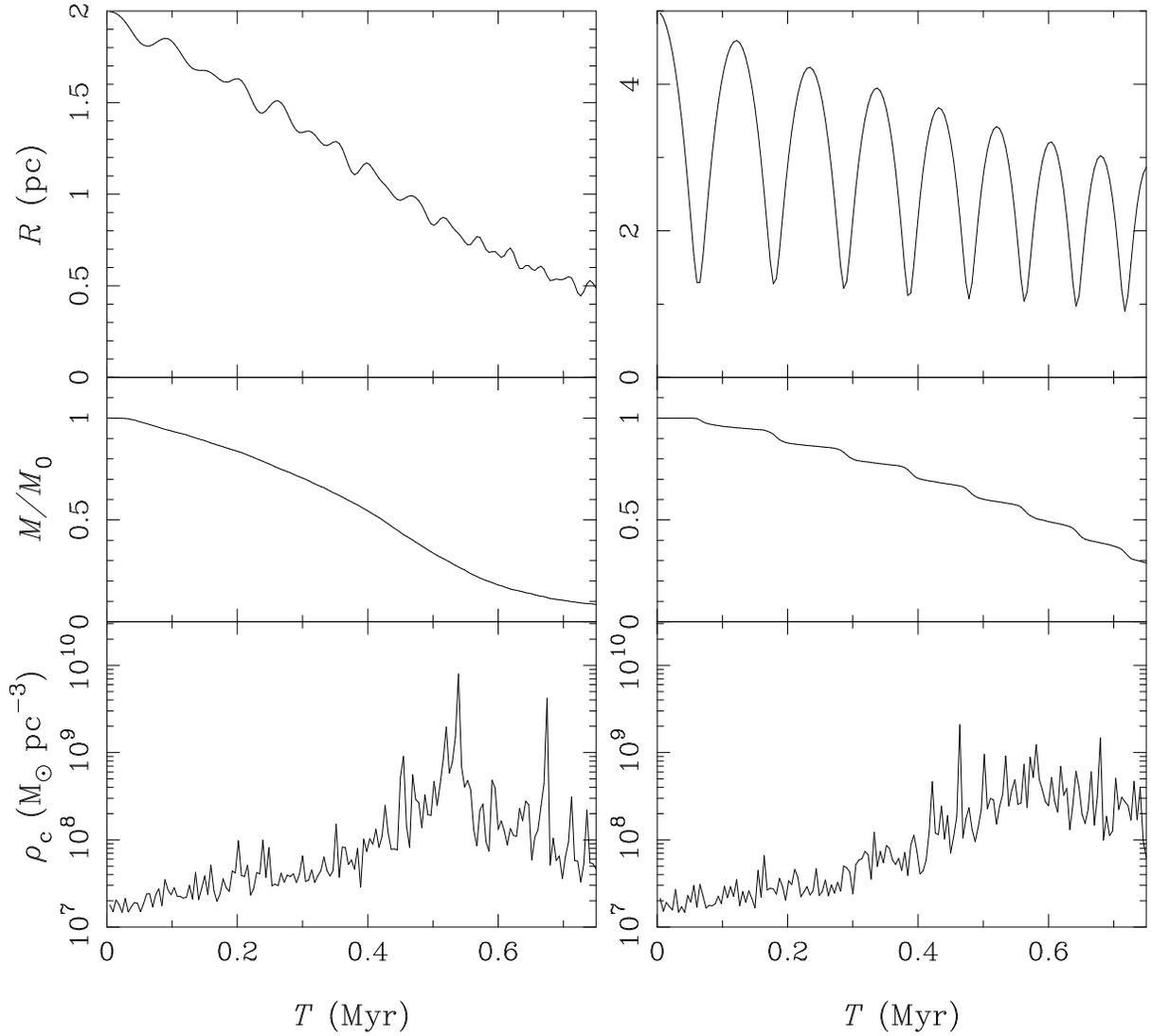

  \begin{center}  
    \FigureFile(160mm,160mm){figure8.eps}
  \end{center}
  \caption{The distance from the GC (top), bound mass (middle), and core
 radius (bottom) of the star clusters plotted as a function of time. 
 The orbits of the star clusters are circular in the left panels and
 eccentric in the right panels.}
  \label{fig:results}
\end{figure*}

Figure \ref{fig:snapshots} show the snapshots from the run in which the
orbit of the star cluster is eccentric.
Figure \ref{fig:results} shows the time evolution of the distance from
the GC, bound mass, and core radius of the star clusters.
In both simulations, core collapse occurs at 0.5 - 0.6
Myr. We obtained the core collapse time, $t_{\rm cc}=0.51$ Myr, from
equation (\ref{eq:t_cc}) and (\ref{eq:t_rh}), where the half-mass radius
of the star cluster, $r_{\rm h}=0.13$ (pc). We adopted $c=0.20$, 
which is suggested by \citep{PM02}. The core
collapse time of our simulations is consistent with the results of the
previous studies.

Figure \ref{fig:err_cir} and \ref{fig:err_ecc} show the total energy
error of the system and the internal energy error of the star clusters.
The internal energy errors are cumulative as in small-$N$ models.
The energies conserve very well.

The total CPU times and the distributions of the CPU time are shown in
table \ref{tb:time}. 
The total CPU time was about 40 hours.
The direct part consumes about half of the CPU time.

\begin{table*}
\begin{center}
\caption{CPU percentage for the test models\label{tb:time}}  
\begin{tabular}{|l|cc|cc|}
\hline
Section of Code & \multicolumn{2}{|c|}{CPU Time (sec)} &
\multicolumn{2}{|c|}{Percentage of CPU Time (\%)}\\
& Circular & Eccentric & Circular & Eccentric\\
\hline \hline
Tree & $5.6\times 10^4$ & $6.3\times 10^4$ & 42.3 & 45.3\\
Direct & $7.5\times 10^4$  & $7.5\times 10^4$ & 56.9 & 53.6\\
The others & $1.1\times 10^3$ & $1.4\times 10^3$ & 0.8 & 1.0\\
\hline
Total & $1.3\times 10^5 \sim 37 (\rm{h})$ 
 & $1.4\times 10^5 \sim 39 ({\rm h})$ & 100.0 & 99.9 \\
\hline
\end{tabular}
\end{center}
\end{table*}

\subsection{Performance Model of the Hybrid Scheme}

\begin{figure}[htbp]
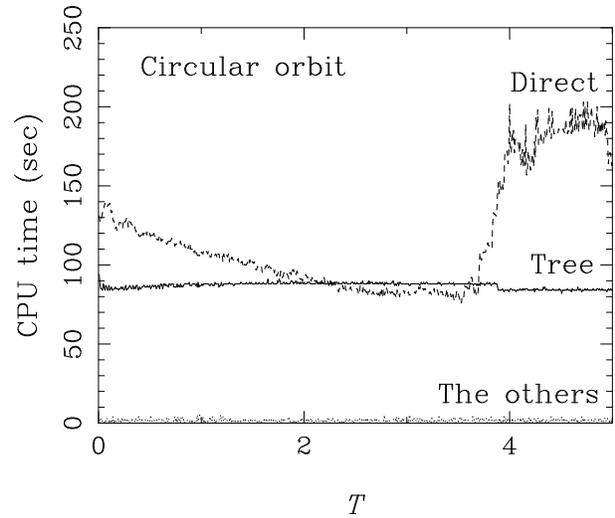

\begin{center}  
 \FigureFile(80mm,50mm){figure11.eps}
 \end{center}
  \caption{CPU time of the direct and tree parts per 4 steps 
($\Delta t = 1/128$) for the circular orbit.
 The solid, dashed, and dotted curves show the CPU time of tree, direct,
 and the others parts, respectively.}
  \label{fig:CPU_time_cir}
\end{figure}

We analyzed the CPU time of each part in detail.
Figure \ref{fig:CPU_time_cir} shows the CPU time per 4 steps for each
part for the run with circular orbit.
Simulation time is represented using the Heggie unit.
We used the parent galaxy to define the Heggie unit, i.e., 
$M_{\rm G}=4E_{\rm G}=1$.
The unit time in the Heggie unit corresponds to 0.15 Myr.
Hereafter we use the Heggie unit for time to discuss the performance of
the hybrid scheme.
The CPU time of the tree part is almost constant throughout the simulation.
In contrast, the cost of the direct part gradually decreases and
suddenly increase after $T \simeq 3.5$.

\begin{figure}[htbp]
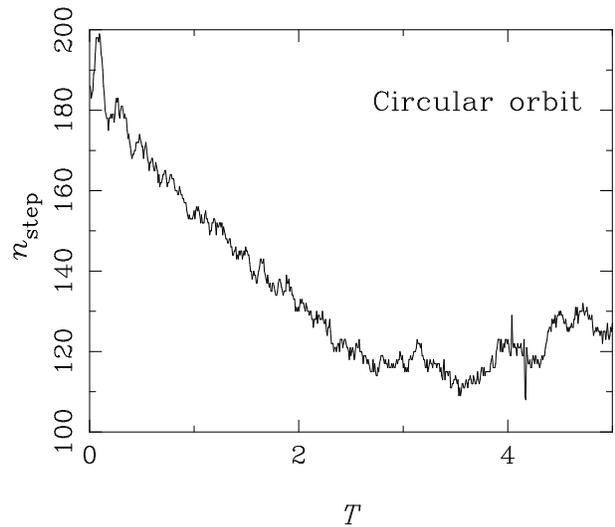

  \begin{center}  
    \FigureFile(80mm,50mm){figure12.eps}
  \end{center}
  \caption{The number of steps of the direct part per particle per 4
 steps (1/128 unit time) for the circular orbit.}
  \label{fig:step_cir}
\end{figure}

\begin{figure}[htbp]
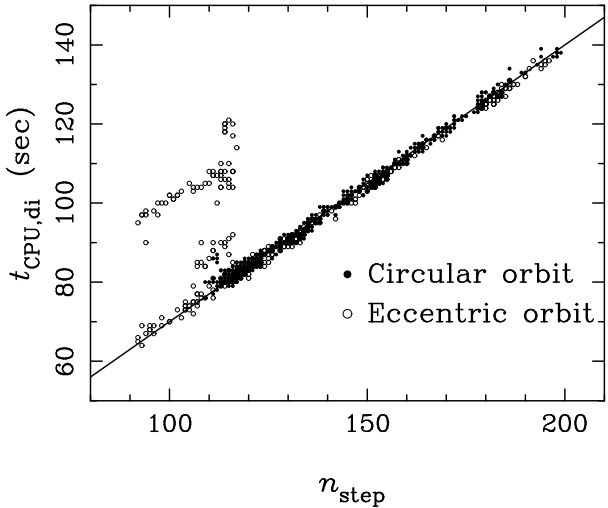

  \begin{center}  
    \FigureFile(80mm,50mm){figure13.eps}
  \end{center}
  \caption{The CPU time of the direct part per 4 steps (1/128 unit time)
 before $T=3.6$. 
 Filled circle and open circle show the results of the circular orbit
 and eccentric orbit. Solid line shows the model of equation
 (\ref{eq:CPU_time}).\label{fig:step}}
\end{figure}

As shown in figure \ref{fig:step}, the CPU time of the direct part is
proportional to the number of the steps for the Hermite scheme,
$n_{\rm step}$.
Figure \ref{fig:step_cir} shows the time evolution of the average number
of timesteps per particle, $n_{\rm step}$. It gradually decrease until
$T=3$, and stays nearly constant. In figure \ref{fig:step}, CPU time
suddenly increases starting at $T=3.6$. This time corresponds to the
time of core collapse, and after that the internal dynamics of the star
cluster is not correctly followed because of the finite softening. In
the discussions below, we consider the behavior of the CPU time before
core collapse.

From these results, we can construct the performance model of the Bridge
code.
The total CPU time for a step, $\Delta t$, can be written by the number
of the tree particles, $N_{\rm tree}$, that of the direct particles, 
$N_{\rm direct}$, and that of steps for Hermite scheme per step,
$n_{\rm step}$.
The cost of tree is proportional to 
$N_{\rm tree}\log N_{\rm tree}\sim N_{\rm tree}$.
The CPU time of the direct part depends on $N_{\rm tree}$ and 
$n_{\rm step}$.
The cost of force calculation is proportional to $N_{\rm direct}^2$ and
the other costs are proportional to $N_{\rm direct}$.
Therefore, the total CPU time per $\Delta t$ is given by
\begin{eqnarray}
T_{\rm CPU} = \alpha N_{\rm tree} + (\beta
 N_{\rm direct} + \gamma N_{\rm direct}^2 ) n_{\rm step},
\label{eq:CPU_time}
\end{eqnarray}
where $\alpha$, $\beta$, and $\gamma$ are constants.
Here, $\alpha$ is almost constant through a simulation, but depends on
$\theta$.
The value of $\gamma$ is determined by the performance of GRAPE. 
Makino et al. (2003) shows that the calculation time on GRAPE per
interaction per particle is expressed as
\begin{eqnarray}
T_{\rm GRAPE} = \frac{1}{9\times 10^7 n_{\rm pipes}}\ ({\rm sec}),
\end{eqnarray} 
where $n_{\rm pipes}$ is the total number of pipelines. With
GRAPE-6A $n_{\rm pipes}=24$, with GRAPE6 $n_{\rm pipes}=192$.
Hence, we estimated $\gamma=4.6\times 10^{-10}$ (sec), for GRAPE-6A,
$\gamma=5.8\times 10^{-11}$ (sec) for GRAPE-6.
From the results of our runs, we estimated the values of the constants,
$\alpha = 1.2\times 10^{-5}$ (sec) for $\theta = 0.75$ and 
$\beta = 6.9 \times 10^{-6}$ (sec).

The number of particles that we need for fully self-consistent $N$-body
star cluster simulation is $N_{\rm G} \sim 2\times 10^6$ for a galaxy 
and $N_{\rm SC} \sim 6.5\times 10^4$ for a star cluster.
In this case, the total CPU time for the Bridge scheme is estimated as
$1.2\times 10^5$ sec $\sim 34$ hours for a simulation with 
$\Delta t = 1/512$ and 5 unit time integration on GRAPE-6. 
Here we used $n_{\rm step}=33$ from the results of our simulations. 
The actual time for such a simulation was 37 - 39 hours
(see table \ref{tb:time}). So our model predicts the CPU time with
$\sim$ 20 \% accuracy.

We can also estimate that for the Hermite scheme.
The CPU time of the Hermite scheme, or the direct scheme, can be
estimated using the second and third terms of equation (\ref{eq:CPU_time}),
where we used $n_{\rm step}=770$ per unit time.
Therefore, the total CPU time is estimated as about $260$ hours per 5
unit times on GRAPE-6 for $N=2\times 10^6$. 
It's about seven times longer than that for the Bridge scheme.

\section{Summary and Discussion}
\subsection{Summary}
We have developed a fast and accurate algorithm, ``the Bridge scheme,''
for fully self-consistent $N$-body simulations of a star cluster moving
in its parent galaxy,  where both are modeled as $N$-body systems.
The Bridge scheme is a hybrid of the tree and direct schemes and
is based on an extension of MVS.
We performed self-consistent $N$-body simulations of a star cluster in a
galaxy and compared the results with the Bridge scheme and that
with the direct scheme (the Hermite scheme). They agreed each other very
well and the energy error was sufficiently small.
We also showed that we can perform a full $N$-body simulation of a
star cluster and a galaxy system with $N_{\rm SC}=65536$ and 
$N_{\rm G}=2\times 10^6$ using our new
scheme more than seven times faster than the direct scheme.

\subsection{Comparison with Tree-based Algorithms}

In previous studies, several tree-based algorithm with block timesteps were
developed. 
\citet{HK89} adopted block timesteps in a tree code.
In their scheme, the tree is reconstructed in each step. 
When the timesteps of particle do not vary so widely, the cost of the
tree reconstruction is not so expensive. However, the cost is very
expensive for star clusters, because star clusters have wide range in
their timesteps.

\citet{MA93} developed a tree-based high-order integration
scheme for collisional systems using block timesteps and multipole (up
to octupole) expansion. 
In this scheme, the tree is reconstructed at the appropriate cell
timesteps determined by the motions of the particles in the cells. 
Instead of reconstructing tree at each step, the moment of each cell is
predicted.
However, the accuracy is limited by the time interval of tree
construction.
If longer timesteps are permitted, the tree evolves during the step and
the errors increase.
If the time interval is short, the cost of tree construction become
large. In addition, their algorithms are difficult to use with GRAPE.

\subsection{Applications to Other Problems}
Our initial motivation for developing the Bridge scheme is to use it for
the problem of a star cluster orbiting in its parent galaxy. However,
it might have much wide application range. For example, if the parent
galaxy has the central massive black hole, it is natural to handle it
and stars near by with the direct scheme, and the rest of the system by
tree. In this case, some particles must move ``tree'' and ``direct''
treatment, but in principle such a code can be developed. Our method can
be applied to any large-$N$ systems in which small part of the system
shows collisional behavior.

\bigskip
The authors thanks Piet Hut for useful comments and the name of the hybrid
scheme, Keigo Nitadori and Ataru Tanikawa for fruitful discussions, and
the referee, Simon  F. Portegies Zwart, for useful comments on the manuscript.
M. F. is financially supported by Research Fellowships of the Japan
Society for the Promotion of Science (JSPS) for Young Scientists.
This research is partially supported by 
the Special Coordination Fund for Promoting Science and Technology
(GRAPE-DR project), Ministry of Education, Culture, Sports, Science and
Technology, Japan.
Part of calculations were done using the
GRAPE system at the Center for Computational Astrophysics (CfCA) of
the National Astronomical Observatory of Japan.


\end{document}